\input harvmac

\input amssym.def
\input amssym.tex
\noblackbox

\def\ZZ{{\Bbb Z}}

\def\CC{{\Bbb C}}

\def\NN{{\Bbb N}}

\def\H{{\cal H}}

\def\M{{\cal M}}

\lref\MatoneBX{
 M.~Matone and R.~Volpato,
 arXiv:math.ag/0506550.
}

\lref\Berkovitsampli{ N.~Berkovits,
 arXiv:hep-th/0503197.
}

\lref\BerkovitsFE{
 N.~Berkovits,
 JHEP {\bf 0004}, 018 (2000)
 [arXiv:hep-th/0001035];
 JHEP {\bf 0009}, 046 (2000)
 [arXiv:hep-th/0006003];
 Int.\ J.\ Mod.\ Phys.\ A {\bf 16}, 801 (2001)
 [arXiv:hep-th/0008145];
 JHEP {\bf 0108}, 026 (2001)
 [arXiv:hep-th/0104247];
 JHEP {\bf 0204}, 037 (2002)
 [arXiv:hep-th/0203248];
 arXiv:hep-th/0209059;
 JHEP {\bf 0409}, 047 (2004)
 [arXiv:hep-th/0406055];
 Comptes Rendus Physique {\bf 6}, 185 (2005)
 [arXiv:hep-th/0410079];
 JHEP {\bf 0503}, 041 (2005)
 [arXiv:hep-th/0411170].
}

\lref\Berkovitss{N.~Berkovits and B.~C.~Vallilo,
 JHEP {\bf 0007}, 015 (2000)
 [arXiv:hep-th/0004171].

 N.~Berkovits and O.~Chandia,
 Phys.\ Lett.\ B {\bf 514}, 394 (2001)
 [arXiv:hep-th/0105149];
 JHEP {\bf 0208}, 040 (2002)
 [arXiv:hep-th/0204121].

N.~Berkovits and P.~S.~Howe,
 Nucl.\ Phys.\ B {\bf 635}, 75 (2002)
 [arXiv:hep-th/0112160].

N.~Berkovits and V.~Pershin,
 JHEP {\bf 0301}, 023 (2003)
 [arXiv:hep-th/0205154].

N.~Berkovits and N.~Seiberg,
 JHEP {\bf 0307}, 010 (2003)
 [arXiv:hep-th/0306226].

N.~Berkovits and D.~Z.~Marchioro,
 JHEP {\bf 0501}, 018 (2005)
 [arXiv:hep-th/0412198].

 N.~Berkovits and N.~Nekrasov,
  arXiv:hep-th/0503075.
  }

\lref\OdaTonin{
 I.~Oda and M.~Tonin,
 Phys.\ Lett.\ B {\bf 520}, 398 (2001)
 [arXiv:hep-th/0109051];
 Phys.\ Lett.\ B {\bf 606}, 218 (2005)
 [arXiv:hep-th/0409052];
  arXiv:hep-th/0505277;
  arXiv:hep-th/0506054.
}

\lref\TrivediMF{
 G.~Trivedi,
 Mod.\ Phys.\ Lett.\ A {\bf 17}, 2239 (2002)
 [arXiv:hep-th/0205217].
}

\lref\ValliloMH{
 B.~C.~Vallilo,
 JHEP {\bf 0212}, 042 (2002)
 [arXiv:hep-th/0210064];
 JHEP {\bf 0403}, 037 (2004)
 [arXiv:hep-th/0307018].

 O.~Chandia and B.~C.~Vallilo,
 JHEP {\bf 0404}, 041 (2004)
 [arXiv:hep-th/0401226].
}

\lref\MukhopadhyayUR{
  P.~Mukhopadhyay,
  arXiv:hep-th/0505157.
}

\lref\SchiappaMK{
 R.~Schiappa and N.~Wyllard,
 arXiv:hep-th/0503123.
}
\lref\MatoneFT{
 M.~Matone, L.~Mazzucato, I.~Oda, D.~Sorokin and M.~Tonin,
 Nucl.\ Phys.\ B {\bf 639}, 182 (2002)
 [arXiv:hep-th/0206104].
}

\lref\AisakaSD{
 Y.~Aisaka and Y.~Kazama,
 JHEP {\bf 0302}, 017 (2003)
 [arXiv:hep-th/0212316];
 JHEP {\bf 0308}, 047 (2003)
 [arXiv:hep-th/0305221];
 JHEP {\bf 0404}, 070 (2004)
 [arXiv:hep-th/0404141];
 arXiv:hep-th/0502208.
}

\lref\GrassiUG{
 P.~A.~Grassi, G.~Policastro, M.~Porrati and P.~Van Nieuwenhuizen,
 JHEP {\bf 0210}, 054 (2002)
 [arXiv:hep-th/0112162].

P.~A.~Grassi, G.~Policastro and P.~van Nieuwenhuizen,
 JHEP {\bf 0211}, 001 (2002)
 [arXiv:hep-th/0202123];
 Adv.\ Theor.\ Math.\ Phys.\ {\bf 7}, 499 (2003)
 arXiv:hep-th/0206216;
 Phys.\ Lett.\ B {\bf 553}, 96 (2003)
 [arXiv:hep-th/0209026];
 Class.\ Quant.\ Grav.\ {\bf 20}, S395 (2003)
 [arXiv:hep-th/0302147];
 Nucl.\ Phys.\ B {\bf 676}, 43 (2004)
 [arXiv:hep-th/0307056];
 arXiv:hep-th/0402122.
}

\lref\AnguelovaSN{
 L.~Anguelova and P.~A.~Grassi,
 JHEP {\bf 0311}, 010 (2003)
 [arXiv:hep-th/0307260].

 P.~A.~Grassi and L.~Tamassia,
 JHEP {\bf 0407}, 071 (2004)
 [arXiv:hep-th/0405072].

 P.~A.~Grassi and P.~van Nieuwenhuizen,
 Phys.\ Lett.\ B {\bf 610}, 129 (2005)
 [arXiv:hep-th/0408007].
}

\lref\CornalbaCU{
 L.~Cornalba, M.~S.~Costa and R.~Schiappa,
 arXiv:hep-th/0209164.
}

\lref\phong{E.~D'Hoker and D.~H.~Phong,
  Phys.\ Lett.\ B {\bf 529}, 241 (2002)
  [arXiv:hep-th/0110247];
  Nucl.\ Phys.\ B {\bf 636}, 3 (2002)
  [arXiv:hep-th/0110283];
  Nucl.\ Phys.\ B {\bf 636}, 61 (2002)
  [arXiv:hep-th/0111016];
  Nucl.\ Phys.\ B {\bf 639}, 129 (2002)
  [arXiv:hep-th/0111040];
  Nucl.\ Phys.\ B {\bf 715}, 91 (2005)
  [arXiv:hep-th/0501196].
  }

\lref\dhok{
 E.~D'Hoker and D.~H.~Phong,
  Nucl.\ Phys.\ B {\bf 715}, 3 (2005)
  [arXiv:hep-th/0501197].
}

\lref\DHokerHT{
 E.~D'Hoker, M.~Gutperle and D.~H.~Phong,
 arXiv:hep-th/0503180.
}

\lref\Vanh{
  M.~B.~Green, H.~h.~Kwon and P.~Vanhove,
  Phys.\ Rev.\ D {\bf 61}, 104010 (2000)
  [arXiv:hep-th/9910055].

  M.~B.~Green and P.~Vanhove,
  Phys.\ Rev.\ D {\bf 61}, 104011 (2000)
  [arXiv:hep-th/9910056].
}

\Title{\vbox{\baselineskip11pt }} {\vbox{ \centerline{Higher Genus
Superstring Amplitudes} \vskip 0.5cm \centerline{From the Geometry
of Moduli Space} \vskip 1pt }}
\smallskip
\centerline{Marco Matone and Roberto Volpato}
\bigskip
\centerline{Dipartimento di Fisica ``G. Galilei'', Istituto
Nazionale di Fisica Nucleare,} \centerline{Universit\`a di Padova,
Via Marzolo, 8 -- 35131 Padova, Italy}

\bigskip
\vskip 0.5cm \noindent

\vskip 1.5cm

\noindent We show that the higher genus 4-point superstring
amplitude is strongly constrained by the geometry of moduli space of
Riemann surfaces. A detailed analysis leads to a natural proposal
which satisfies several conditions. The result is based on the
recently derived Siegel induced metric on the moduli space of
Riemann surfaces and on combinatorial products of determinants of
holomorphic abelian differentials.

\Date{June 2005}
%
\baselineskip14pt

%
%


\newsec{Introduction}

One basic problem in superstring theory concerns the explicit
evaluation of the higher genus amplitudes. A fundamental step in
such a direction has been done by D'Hoker and Phong that in a series
of remarkable papers \phong\dhok\DHokerHT, provided an explicit
gauge slice independent formulation of the NSR superstring at genus
two. Their analysis also provides relevant suggestions for the
higher genus generalization. The fact that the quantum string is
described in terms of classical geometry of the moduli space
suggests trying to investigate the structure of the higher genus
amplitudes from purely algebraic-geometrical considerations.

In recent years, a new covariant formulation of the superstring has
been proposed by Berkovits
\refs{\Berkovitsampli\BerkovitsFE\Berkovitss\OdaTonin\MatoneFT\GrassiUG\AnguelovaSN\ValliloMH\TrivediMF\SchiappaMK\AisakaSD\CornalbaCU-\MukhopadhyayUR}.
In this respect we observe that the analysis of (super)strings leads
to consider basic properties of the moduli space of Riemann
surfaces. Below we will comment on a possible algebraic-geometrical
interpretation of the pure spinor condition. Such a condition is a
bilinear relation on differentials, whose weight is changed after
the twisting, depending on the superstring critical dimension. Known
classical relations among differentials concern, $e.g.$, the
characterization of a basis of holomorphic quadratic differentials,
out of the $g(g+1)/2$ one may get by bilinear combinations of the
holomorphic one-differentials (Schottky problem). Other relations
among holomorphic differentials which, like the pure spinor
condition, involves the critical dimension, is provided by the
Mumford isomorphism.

In this paper we focus on the higher genus 4-point superstring amplitude. In the case of genus 2, this problem has been first
considered by Iengo and Zhu \ref\IengoZK{
  R.~Iengo and C.~J.~Zhu,
  Phys.\ Lett.\ B {\bf 212}, 309 (1988);
  %
  Phys.\ Lett.\ B {\bf 212}, 313 (1988).
}. Following the methods of \phong, an explicit formula, still gauge slice dependent, has been derived in 
\ref\Zheng{
  Z.~J.~Zheng, J.~B.~Wu and C.~J.~Zhu,
  %
  Phys.\ Lett.\ B {\bf 559}, 89 (2003)
  [arXiv:hep-th/0212191];
  Nucl.\ Phys.\ B {\bf 663}, 79 (2003)
  [arXiv:hep-th/0212198];
  %
  Nucl.\ Phys.\ B {\bf 663}, 95 (2003)
  [arXiv:hep-th/0212219].
}. The basic problem of a fully gauge slice independent formulation has been finally resolved in \dhok\ \eqn\tauswa{ A_4^{\rm
2-loop} = B_2 \, \int_{{\cal M}_2}{ |\bigwedge_{i\le j}^2 {\rm d} \Omega_{ij}|^2   \over (\det  {\rm Im} \, \Omega)^3 }\,
\int_{\Sigma^4} {|{\cal Y}_S|^2\over (\det  {\rm Im} \, \Omega)^2 } \, e^{-\sum_{i<j}k_i\cdot k_j\,G(z_i,z_j)}\ , } where
$${\cal Y}_S=(k_1-k_2)\cdot(k_3-k_4)\Delta (z_1,z_2) \Delta
(z_3,z_4) + 2\leftrightarrow 3 + 2\leftrightarrow 4 \ ,
$$
$$
G(z,w):= - \ln |E(z,w)|^2 + 2 \pi {\rm Im}\int_z^w\omega_i ({\rm
Im}\Omega)^{-1}_{ij}  {\rm Im} \int_z^w \omega_j \ ,
$$
and
$$\Delta(z,w):=\omega_1(z)\omega_2(w)-\omega_2(z)\omega_1(w)\ ,
$$ is the basic holomorphic antisymmetric bi-differential which plays
the role of building block for the genus 2 superstring amplitudes. A
preliminary investigation concerning the extension to higher loop of
\tauswa\ has been
considered in 
\ref\ZhuJN{
  C.~J.~Zhu,
  arXiv:hep-th/0503001.
  }. A similar formula has been recently derived by Berkovits in his
  pure spinor formalism \Berkovitsampli.

To find the higher genus extension of such an amplitude requires
considering two problems

\vskip 0.333cm

\item{1.} Find the higher genus version of the modular invariant measure
$${|\bigwedge_{i\le j}^2 {\rm d} \Omega_{ij}|^2   \over
(\det  {\rm Im} \, \Omega)^3 }\ . $$ Whereas in the case of genus 3
the analog of such a volume form can still be used, except for the
hyperelliptic locus where by N\"other theorem the bilinears
$\omega_i(z)\omega_j(z)$ span a $(2g-1)$-dimensional subspace of
$H^0(\Sigma,K^2)$, in the case $g>3$ the
${1\over2}g(g+1)$-dimensional space of symmetric $g\times g$ matrix
with positive definite imaginary part, should satisfy conditions
leading to a subspace parametrizing the moduli space. \vskip 0.333cm

\item{2.} The other problem with the higher genus extension of 4-point
amplitude concerns the generalization of
$$
 \int_{\Sigma^4} {|{\cal Y}_S|^2\over (\det  {\rm Im}
 \, \Omega)^2 } \, e^{-\sum_{i<j}k_i\cdot k_j\,G(z_i,z_j)}\ . $$
On the other hand, whereas $G(z_i,z_j)$ is canonically defined for
any $g$, there is not an obvious extension of ${\cal Y}_S$. However,
we will see that we are essentially forced to still use the
determinants of the holomorphic one-differentials as building block
for constructing the ${\cal Y}_S$. This implies that one has to
introduce $g-2$ more points for each determinant. Since the
determinants come in pair, we need to introduce $2g-4$ points. We
then are lead to analyze the structure of the possible permutations
between such points and then integrating on $\Sigma^{2g-4}$.

\vskip 0.333cm

\noindent Then our proposal is \eqn\aquattro{ A_4^{g\rm-loop} =
B_g \, \int_{\M_g}  |\det g|^{1/2}{\rm
d}\,\Xi_1\wedge\ldots\wedge {\rm d}\,\bar\Xi_{3g-3} {\cal F}_g(k_i)\
, } where $g$ is the Siegel induced metric derived in \MatoneBX,
${\rm d}\,\Xi_i$, $i=1,\ldots,3g-3$ are the moduli induced by the
constraints on $d\Omega_{ij}$ \MatoneBX. Furthermore,
$$
{\cal F}_g(k_i)=\int_{\Sigma^{2g}} {|{\cal Y}_S|^2\over(\det  {\rm
Im} \, \Omega)^2 } \, e^{-\sum_{i<j}k_i\cdot k_j\,G(z_i,z_j)}\  ,
$$
and \eqn\ipsilon{{\cal Y}_S=(k_1-k_2)\cdot(k_3-k_4)\H_{12,34}+
2\leftrightarrow 3 + 2\leftrightarrow 4 \ , } where $\H_{12,34}$ is
a sum on a set of permutations of $2g-4$ points of terms such as
$$
X_{12}(w_{1\rightarrow g-2})X_{34}(w_{g-1\rightarrow 2g-4}) \ ,
$$
where $$X_{ij}(w_{k\rightarrow m}) :=\det
\omega(z_i,z_j,w_k,\ldots,w_m)\ .
$$
Defining ${\cal Y}_S$ in terms of determinants of holomorphic
differentials and then integrating on the $2g-4$ variables
$\{w_k\}$, provides a way to get modular invariant linear
combinations, of terms such as
$|\Delta_{ij}(z_1,z_2)\Delta_{kl}(z_3,z_4)|^2$, where
$\Delta_{ij}(z_k,z_l):=\omega_i(z_k)\omega_j(z_l)-\omega_j(z_k)\omega_i(z_l)$,
with the coefficients given by bilinears in the minors of order
$g-2$ of ${\rm Im}\,\Omega$ divided by $(\det  {\rm Im} \,
\Omega)^2$. For example, as we will see, integrating over
$w_1,\ldots,w_{g-2}$ the simplest product, we have
$${1\over2^{g-2}(g-2)!}\int_{\Sigma^{g-2}}\prod_k|dw_k|^2
\det \omega(z_1,z_2,w_{1},\ldots,w_{g-2})\det
\bar\omega(z_3,z_4,w_1,\ldots,w_{g-2})=
$$
$$\sum_{i<j\atop
m<n}(-1)^{i+j+m+n}\Delta_{ij}(z_1,z_2)\bar\Delta_{mn}(z_3,z_4)
\det_{k\neq i,j\atop l\neq m,n} {\rm Im}\,\Omega_{kl}\ .
$$
{}From a geometrical point of view, our investigation corresponds to
find the higher genus analog of \tauswa. Thus, besides modular
invariance, one should care about the absence of ambiguities, such
as the dependence on the choice of arbitrary points in the higher
loop generalization of \tauswa. Using the determinants of the
standard basis of the holomorphic differentials and then integrating
over the additional $2g-4$ points removes possible ambiguities and
guarantees modular invariance. Therefore, the main properties of the
amplitude we derived are

\vskip 0.333cm

\item{a.}{\it Manifestly ambiguity free.}

\item{b.}{\it Modular invariance.}

\item{c.}{\it Reproduces the $g=2$ expression.}

\vskip 0.333cm

Among the possible permutations of the points $\{w_k\}$, we will
focus on two particularly simple cases. The simplest one corresponds
to consider \eqn\ball{ {\cal H}_{12,34}=X_{12}(w_{1\rightarrow
g-2})X_{34}(w_{g-1\rightarrow 2g-4})+ X_{34}(w_{1\rightarrow
g-2})X_{12}(w_{g-1\rightarrow 2g-4})\ , } so that \ipsilon\
guarantees the invariance of ${\cal Y}_S$ under the simultaneous
exchange of $k_i\leftrightarrow k_j$, $z_i\leftrightarrow z_j$.
Since this is also the property of the exponential in \aquattro, the
resulting amplitude is symmetric under exchange of the momenta.

It is interesting to understand the general structure of ${\cal
H}_{ij,kl}$ once one considers different choices of the possible
permutations, an issue considered in detail in the paper. Since by
construction the amplitude is invariant under permutation of the
external momenta, it is clear that different choices of permutations
of the $\{w_k\}$ will always lead to an expression of
$\int_{\Sigma^{2g}}|{\cal Y}_S|^2$ proportional to
$(s^2+t^2+u^2)(\det{\rm Im}\,\Omega)^2$, where the term $(\det{\rm
Im}\,\Omega)^2$ follows by the properties of $|{\cal Y}_S|^2$ under
modular transformations. Nevertheless, because of the exponential of
the Green-functions, different choices would in principle lead to
different expressions for the amplitude.

The fact that modular invariance essentially leads to introduce
determinants of the canonical basis of holomorphic one-differentials
differentials $\omega_k$, suggests that the pair of extra $g-2$
differentials $\omega_k$ is due to zero modes. In the standard
approach such zero modes are related to the spin structures of the
$\beta-\gamma$ system, while, interestingly enough, in the Berkovits
approach the zero modes are related to other elementary fields.

We note that in \Berkovitsampli, besides the derivation of the
4-point $g=2$ superstring amplitude, Berkovits also notes that
preliminary calculations in his formulation indicate that the
momentum dependence of the higher genus 4-point amplitude is the
same of the one in genus two, in agreement with the suggestion by
Zhu in \ZhuJN. On the other hand, on general grounds, it is
difficult to understand the mechanism for which terms giving
contributions to $D^4R^4$, having the same analytic general
structure of the genus two case, may be suppressed. However, such a
possibility cannot be excluded, for example by summing on
antisymmetric permutations of the $\{w_k\}$, an operation which is
reminiscent of the summation over the odd spin structures, one may
get a vanishing result. Nevertheless it remains the problem of
finding an analytic structure for the amplitude which takes place
only at genus greater than two. Another possibility is that for
$g>2$ the hyperelliptic contribution, that we do not consider here,
may lead to some cancellation mechanism. On the other hand, the
inclusion of the hyperelliptic locus corresponds to a generalization
of \aquattro\ involving essentially only the measure. Since the
hyperelliptic locus has codimension $g-2$, also in this case one
should understand the structure of such a cancellation. It would be
interesting to fully understand the basic role of $S$-duality for
the higher genus amplitudes along the lines investigated in \Vanh.

The paper is organized as follows. In section 2 we first consider
the problems which arise in choosing additional points in defining
the $g\times g$ matrices $\omega_i(x_j)$. We will also comment on
some properties satisfied by the basis of holomorphic differentials
which are related both to the problem of the measure on ${\cal M}_g$
(Schottky problem), and to some speculation on the possible
geometrical understanding of the pure spinor condition. Next, we
will shortly review the measure on the (coarse) moduli space
corresponding to the generalization to higher genus of the $g=2$
measure.

In section 3 we will start by constructing ${\cal Y}_S$ by means of
a suitable combination of products of determinants. We then will
consider in some detail the underlying combinatorial structure which
arises in considering the permutations of the additional points
$\{w_k\}$. As we will see, the kind of analysis suggests further
developments which may be of mathematical interest as well.

Finally, section 4 is devoted to conclusions and suggestions for future investigations.

\newsec{Geometry of ${\cal M}_g$ and the modular invariant metric}

There are two main questions in considering the higher genus
generalization of the 4-particle amplitude, the measure on the
moduli space and the higher genus generalization of the holomorphic
bi-differential $\Delta(z,w)$. Let us first concentrate on the
latter. On general grounds, we should look at the simplest possible
generalization. This would suggest that even in higher genus
\eqn\mayexpect{ \Delta(z,w)=\det \omega_i(z_j) \ , } with $z_1=z$
and $z_2=w$. However, such an expression would require choosing the
$g-2$ points $z_i$, $i=3,\ldots,g$, that, in order to avoid
ambiguities in defining the amplitude,  should be chosen in a
canonical way. Also note that a naive integration along cycles of
$\Sigma$ on each of the $g-2$ points would spoil the modular
properties of $\Delta(z_i,z_j)$. At first sight, however, there is
no a canonical way to fix such a set of points. In particular, we
cannot expect that this set would correspond to the zeroes of some
differential. The reason is that the degree of a
$\lambda$-differential is $2\lambda(g-1)$, so that fixing a set of
canonical points should require a number of points multiple of
$g-1$. A typical example concerns the $g-1$ zeroes of the
holomorphic one-differentials constructed in terms of the
$\theta$-function with odd spin structure. This zero set depends on
the period matrix and on the particular odd spin structure, so it
reduces the freedom in choosing the points to the choice of the
particular odd spin structure.\foot{For hyperelliptic Riemann
surfaces of $g=3$ one can use a $\theta$-function with even
characteristic.}

There are several structures that start emerging at genus three. One
concerns the dimensionality of the locus where the $\theta$-function
vanishes identically. To understand when this happens, note that
linear independence of the $g$ holomorphic abelian differentials may
fail at specific points. This means that there are divisors for
which the rows of $\omega_i(z_j)$ are linearly dependent. Writing
$\det \omega_i(z_j)$ in terms of $\theta$-functions and prime forms,
one sees that such a locus of {\it super zeroes} is $g-2$
dimensional.

Another place where appears a dependence on $g-2$, concerns the
systematic construction by Petri of a basis for $H^0(\Sigma,K^2)$ by
means of bilinears of a suitable basis of $H^0(\Sigma,K)$
\ref\ottimo{E. Arbarello, M. Cornalba, P.A. Griffiths and J. Harris,
``Geometry of Algebraic Curves", Vol.1. Berlin, Heidelberg,
New-York, Tokyo,
1985.} (see also \MatoneBX\ and 
\ref\ManinGX{
  Y.~I.~Manin,
  Phys.\ Lett.\ B {\bf 172}, 184 (1986).
  A.~A.~Beilinson and Y.~I.~Manin,
  Commun.\ Math.\ Phys.\ {\bf 107}, 359 (1986).
} for related constructions). In turn, this is related to the
mentioned fact that the codimension of the hyperelliptic locus in
the moduli space of Riemann surfaces is just $g-2$.

\subsec{Critical dimension and Mumford isomorphism.}

The geometry of the moduli space is deeply connected with the
field-theoretical formulation of string theory. In particular, there
are properties, such as the critical dimension, which are directly
connected to basic relations satisfied by the holomorphic
differentials. For example, in the $g>3$ non-hyperelliptic case,
constructing a basis of holomorphic quadratic differentials in terms
of bilinear combinations of the canonical basis of holomorphic
one-differentials $\omega_i$ is the Schottky problem. A twisting in
the field content may be connected to such relations. There is
another relevant relationship between holomorphic one- and
two-differentials, the Mumford isomorphism
$$
K\simeq E^{13} \ ,
$$
where $E$ is the Hodge line bundle and $K$ the determinant line
bundle. Schottky problem and Mumford isomorphism involve strictly
related structures. On the other hand, the Mumford isomorphism
provides an algebraic-geometrical understanding of the bosonic
critical dimension \ManinGX\ref\BelavinCY{
  A.~A.~Belavin and V.~G.~Knizhnik,
  Phys.\ Lett.\ B {\bf 168}, 201 (1986).

  E.~P.~Verlinde and H.~L.~Verlinde,
  Nucl.\ Phys.\ B {\bf 288}, 357 (1987).

  E.~D'Hoker and D.~H.~Phong,
  Rev.\ Mod.\ Phys.\  {\bf 60}, 917 (1988).
}. We also note that such issues are strictly related to the problem
of classifying the higher genus modular forms, and so to the problem
of writing the Mumford form in terms of $\theta$-constants and
related objects (see
$e.g.$ 
\ref\MorozovDA{A.~Morozov,
  Phys.\ Lett.\ B {\bf 184}, 171 (1987).
  }).

There is another relationship involving differentials on Riemann
surfaces and critical dimension. This is the pure spinor condition
\eqn\purespinor{
\lambda_\alpha\gamma^m_{\;\;\alpha\beta}\lambda_\beta=0\ , \qquad
\bar\lambda_\alpha\gamma^m_{\;\;\alpha\beta}\bar\lambda_\beta=0\ .}
Here $m$ runs from 0 to 9 and $\gamma^m_{\alpha\beta}$ are
$16\times16$ matrices which are off-diagonal blocks of the
$32\times32$ ten-dimensional $\Gamma$-matrices and satisfy
$$
\gamma^m_{\;\;\alpha\beta}
\gamma^{n\,{\beta\gamma}}=2\eta^{mn}\delta_\alpha^\gamma \ .$$
Originally the $\lambda$'s and $\bar\lambda$'s are considered
$({1\over2},0)$-differentials and $(0,{1\over2})$-differentials
respectively. After twisting the $\lambda$'s become
$(0,0)$-differentials and $\bar\lambda$'s $(0,1)$-differentials.

Note that bilinear relations among holomorphic abelian differentials
may lead to conditions on scalars and $1/2$-differentials. This
suggests that the pure spinor condition \purespinor\ may be related
to the Schottky problem. In this respect, it is worth noticing that
the Schottky problem itself is related to the uniformization problem
via the Liouville equation
\ref\MatoneKR{
  M.~Matone,
  Lett.\ Math.\ Phys.\ {\bf 33}, 75 (1995)
  [arXiv:hep-th/9310066].
}.

Let us also notice that in the pure spinor formalism the $b$-ghost
is a composite field. It would be interesting to understand whether
there exists a suitable combination of zero modes of elementary
fields that may suitably combine to build the volume form on ${\cal
M}_g$ made of quadratic holomorphic differentials.

\subsec{The induced measure}

We now consider the problem of defining the higher genus version of
the measure on the moduli space in \tauswa. Let $\hat{\cal M}_g$ be
the locus of moduli space of compact non-hyperelliptic Riemann
surfaces of genus $g\geq 4$. The restriction on $\hat{\cal M}_g$ of
the measure on the Siegel upper half-space has been derived in
\MatoneBX. In the following we briefly review the results of this
paper.

Let $\Sigma$ be a compact non-hyperelliptic Riemann surface of genus
$g\ge 4$ and $\{\omega_1,\ldots,\omega_g\}$ the canonical basis of
$H^0(\Sigma,K)$, with $K$ the canonical line bundle of $\Sigma$. Let
$p_1,\ldots,p_g$ be a set of points in $\Sigma$ such that
$$\det \omega(p_1,\ldots,p_g)\ne 0\ ,$$ where
$\det \omega(p_1,\ldots,p_g):=\det \omega_i(p_j)$. Then
\eqn\newbasis{\sigma_i(z):={\det \omega(p_1,p_2,\ldots,p_{i-1},
z,p_{i+1},\ldots,p_g)\over \det \omega(p_1,\ldots,p_g)}\ ,}
$i=1,\ldots,g$, is a basis of $H^0(\Sigma,K)$.

\vskip 5pt

\noindent We introduce the $g(g+1)/2$-dimensional vector $v$
$$v_k:=\left\{\matrix{\sigma_k^2\ ,\hfill\qquad\hfill &k=1,\ldots,g\
,\hfill\cr &\cr \sigma_{i+j}\sigma_j\ ,\hfill\qquad \hfill
&k=i+j(2g-j+1)/2\ \ ,\hfill}\right .$$ where $j=1,\ldots,g-1$,
$i=1,\ldots,g-j$. If we assume that the divisors of $(\sigma_i)$
consist of distinct points, then $\{v_j\}:=\{v_1,\ldots,v_{3g-3}\}$
is a modular invariant basis of $H^0(\Sigma,K^2)$, the space of the
holomorphic quadratic differentials on $\Sigma$.

Let $W(P)$ be the Wronskian $W(v_1,\ldots,v_{3g-3})(P)$ of the basis
$\{v_j\}$ at a generic point $P\in \Sigma$ and $\hat{W}_{k,ij}(P)$
be the Wronskian
$W(v_1,\ldots,v_{k-1},\omega_i\omega_j,v_{k+1},\ldots,v_{3g-3})(P)$
at $P$. We have

\vskip 10pt

\proclaim Theorem 1.
\eqn\teor{\omega_i(z)\omega_j(z)=\sum_{k=1}^{3g-3}{\hat{W}_{k,ij}\over
W}v_k(z)\ ,} where the ratio $\hat{W}_{k,ij}/W$ does not depend on
the point $P$.

\vskip 10pt

\noindent The line element $ds$ on the Siegel upper half-space

$$ds^2:=\Tr\, [({\rm Im}\, \Omega)^{-1}d\Omega ({\rm Im}\,
\Omega)^{-1}d\bar\Omega] \ ,$$ defines the volume element
$${ |\bigwedge_{i\le j}^g {\rm d} \Omega_{ij}|^2\over
(\det {\rm Im}\, \Omega)^{g+1} } \ . $$

\vskip 6pt \noindent Let $k$ be the Kodaira-Spencer map identifying
the quadratic differentials on $\Sigma$ with the fiber of the
cotangent of the Teichm\"uller space at the point representing
$\Sigma$. We have $$k(\omega_i\omega_j)={(2\pi i)}^{-1}d\Omega_{ij}
\ .$$ Let us set
$$
d\,\Xi_i:= 2\pi i\,k(v_i)\ ,\qquad i=1,\ldots,3g-3 \ .
$$
By Eq. \teor\ we have $$d\Omega_{ij}=\sum_{k=1}^{3g-3}
{\hat W_{k,ij}\over W}d\,\Xi_k \ ,$$ $i,j=1,\ldots,g$.

\vskip 10pt

\proclaim Theorem 2. The line element $ds_{|\hat\M_g}$ on $\hat{\cal
M}_g$ induced by the Siegel metric is $$ds_{|\hat\M_g}^2:=
\sum_{i,j=1}^{3g-3}g_{ij}d\,\Xi_id\,\bar\Xi_j \ ,$$ where
$$g_{ij}:= |W|^{-2}\Tr\, [({\rm Im}\, \Omega)^{-1}\hat W_i
 ({\rm Im}\, \Omega)^{-1}\bar{\hat W_j} ] \ .$$

\vskip 10pt

\noindent It follows that the Siegel induced modular invariant
volume form on $\hat{\cal M}_g$ is $$d\nu:=|\det
g|^{1\over2}d\,\Xi_1\wedge\ldots \wedge d\,\bar \Xi_{3g-3} \ .$$

\newsec{The modular function ${\cal Y}_S$.}

In the following we will define the higher genus version of the
genus 2 formula for ${\cal Y}_S$. In particular, we generalize the
building blocks $\Delta(z_1,z_2)\Delta(z_3,z_4)$, used to construct
${\cal Y}_S$ in the genus two case, as a product of a pair of
determinants of holomorphic one-differentials on the Riemann surface
$\Sigma$. We will be forced to introduce $2g-4$ arbitrary points,
that will be integrated over $\Sigma^{2g-4}$ in order to obtain an
unambiguous result. The formula will result to be modular invariant
and to reproduce the genus 2 expression of the 4-point superstring
amplitude.

At genus 3, the above requirements fix, up to a normalization
constant, the form of ${\cal Y}_S$. At higher genus we have to deal
with several possibilities. Although we will consider all these
possibilities, we will show that there exists a simple formula which
seems a natural generalization of the genus two case.

\subsec{Definitions and properties}

Let us define
$$\matrix{&X(w_{k\rightarrow n})\hfill &:=
\det \omega(w_k,\ldots,w_n)\ ,\hfill\cr\cr &X_{ij}(w_{k\rightarrow
m})\hfill &:= \det \omega(z_i,z_j,w_k,\ldots,w_m)\ ,\hfill\cr\cr
&X(w_{\rho(k\rightarrow n)})\hfill &:= \det
\omega(w_{\rho(k)},\ldots,w_{\rho(n)}) \ ,\hfill\cr\cr
&X_{ij}(w_{\rho(k\rightarrow m)})\hfill &:= \det
\omega(z_i,z_j,w_{\rho(k)},\ldots,w_{\rho(m)})\ ,\hfill}$$ for all
$k,n,m\in{\NN}$, with $n-k+1=g$ and $m-k+1=g-2$, and set
\eqn\Hijkl{\H_{ij,kl}:=\sum_{\rho\in
B_{2g-4}}C(\rho)X_{ij}(w_{\rho(1\rightarrow
g-2)})X_{kl}(w_{\rho(g-1\rightarrow 2g-4)})\ ,} where
$w_1,\ldots,w_{2g-4}$ are points in $\Sigma$, $C(\rho)$ are
coefficients in ${\CC}$ and $B_{2g-4}$ is a subset, that we will
define below, of the group of permutations of $\{1,2,\ldots,2g-4\}$.
Essentially, we want the elements in $B_{2g-4}$ to be in one to one
correspondence with all the different ways of choosing $g-2$ points
out of the $2g-4$ points $\{w_i\}$. Therefore, we require that for
any pair $\rho,\sigma\in B_{2g-4}$, $\rho\neq\sigma$ implies
$\rho(\{1,2,\ldots,g-2\})\neq\sigma(\{1,2,\ldots,g-2\})$. There is
not a unique way to choose the elements in $B_{2g-4}$; different
choices can give some minus sign that can be reabsorbed by a
redefinition of the coefficients $C(\rho)$.

We also impose the following constraint on $\{C\}$: if a pair
$\rho,\sigma\in B_{2g-4}$, satisfies
$\rho(\{1,2,\ldots,g-2\})=\sigma(\{g-1,g,\ldots,2g-4\})$ then
$C(\rho)=C(\sigma)$. This ensures that $\H_{ij,kl}=\H_{kl,ij}$. One
can verify that, at genus $g$, we have to fix $(2g-4)!/2[(g-2)!]^2$
independent coefficients $C$. Each of them appears twice in \Hijkl.
We can avoid this redundancy using the equivalent definition
$$\H_{ij,kl}:=
$$
$$\sum_{\rho\in B'_{2g-4}}C(\rho) [X_{ij}(w_{\rho(1\rightarrow
g-2)})X_{kl}(w_{\rho(g-1\rightarrow 2g-4)})+
X_{kl}(w_{\rho(1\rightarrow g-2)})X_{ij}(w_{\rho(g-1\rightarrow
2g-4)}))]\ ,$$ where $\rho\in B'_{2g-4}\subset B_{2g-4}$ satisfies
the condition $\rho(1)\leq g-2$. We define
$${\cal
Y}_S:=(k_1-k_2)\cdot(k_3-k_4)\H_{12,34} + 2\leftrightarrow 3+
2\leftrightarrow 4 \ ,
$$
and propose that the integral in $\Sigma^4$ is now replaced by the
integral over $\Sigma^{2g}$ \eqn\amplitude{ {\cal
F}_g(k_i):=\int_{\Sigma^{2g}} {|{\cal Y}_S|^2\over (\det  {\rm Im}
\, \Omega)^2 } \, e^{-\sum_{i<j}k_i\cdot k_j\,G(z_i,z_j)}\ , }
\vskip 10pt \noindent where $G(z_1,z_2)=-\ln|E(z_1,z_2)|^2+2\pi{\rm
Im}\int_{z_1}^{z_2} \omega_i\,({\rm Im\,\Omega})^{-1}_{ij}\,{\rm
Im}\int_{z_1}^{z_2}\omega_j$. Under a modular transformation
$\tilde\Omega=(A\Omega +B)\cdot(C\Omega+D)^{-1}$, we have \vskip 3pt
$$
\det  {\rm Im}\, \tilde\Omega=|\det (C\Omega+D)|^{-2}\det  {\rm
Im}\, \tilde\Omega\ ,$$ $$|\tilde{\cal Y}_S|^2=|\det
(C\Omega+D)|^{-4}|{\cal Y}_S|^2\ ,
$$
so that ${\cal F}_g(k_i)$ is modular invariant.

Performing the integration over the $2g-4$ variables $\{w_k\}$ in
\amplitude\ results in a linear combination of terms such as
$|\Delta_{ij}(z_1,z_2)\Delta_{kl}(z_3,z_4)|^2$, times $
e^{-\sum_{i<j}k_i\cdot k_j\,G(z_i,z_j)}$, where \vskip 6pt
$$
\Delta_{ij}(z_k,z_l):=\omega_i(z_k)\omega_j(z_l)-\omega_j(z_k)\omega_i(z_l) \ ,
$$
\vskip 6pt \noindent with the coefficients given by bilinears in the
minors of order $g-2$ of ${\rm Im}\,\Omega$ divided by $(\det  {\rm
Im} \, \Omega)^2$. To see this, first observe that since the
determinant of a $n\times n$ matrix can be written as
$$\det a=\epsilon_{i_1\ldots i_n}a_{1i_1}\ldots a_{ni_n}={1\over
n!}\epsilon_{i_1\ldots i_n}\epsilon_{j_1\ldots j_n}a_{i_1j_1}\ldots
a_{i_nj_n}\ ,$$ we have \vskip 6pt
$$
\det\int_\Sigma|dz|^2\omega_i\bar \omega_j={1\over g!}
\int_{\Sigma^g}\prod_k|dz_k|^2 \epsilon_{i_1\ldots
i_g}\epsilon_{j_1\ldots j_g}\omega_{i_1}(z_1)\bar
\omega_{j_1}(z_1)\ldots\omega_{i_g}(z_g)\bar \omega_{j_g}(z_g)\ ,
$$
\vskip 6pt \noindent so that, considering the index of the
integration variables as matrix index, \eqn\bbbb{ {1\over
g!}\int_{\Sigma^g}\prod_k|dz_k|^2 |\det
\omega_i(z_j)|^2=\det\int_{\Sigma} \omega_i\wedge\bar
\omega_j=2^g\det{\rm Im}\,\Omega_{ij}\ , } where \eqn\omom{
\int_{\Sigma}|dz|^2\omega_i\bar\omega_j:=
-i\int_{\Sigma}\omega_i\wedge\bar\omega_j=2 {\rm Im}\,\Omega_{ij}\ .
} \vskip 6pt \noindent In order to explicitly perform the
integration of the $2g-4$ variables $\{w_i\}$ in \amplitude, we
factorize the dependence on the $\{z_i\}$ in ${\cal H}_{ij,kl}$. To
this end, note that \vskip 6pt
$$
\epsilon_{i_1\ldots i_g}=(-1)^{i_1+i_2+1}\epsilon_{i_1i_2}
\epsilon^{(i_1i_2)}_{i_3\ldots i_g}\ ,
$$
\vskip 6pt \noindent where the value of the superscripts $i_1$ and
$i_2$ are excluded from the range of values of $i_{k>2}$ and
$\epsilon_{i_1i_2}=1$, $\epsilon^{(i_1i_2)}_{i_3\ldots i_g}=1$ if
$i_1<i_2$ and $i_3<\ldots<i_g$, respectively. Therefore, we have
\eqn\Atausw{ \det
\omega(z_1,z_2,w_{1},\ldots,w_{g-2})=\sum_{i<j}(-1)^{i+j+1}
[\omega_i(z_1)\omega_j(z_2)-\omega_j(z_1)\omega_i(z_2)]\det_{k\neq
i,j}\omega_k(w_l)\ , } where $l=1,\ldots,g-2$. By means of such a
factorization, we may perform the integration on the $2g-4$
variables $\{w_i\}$ in \amplitude. For example, by \Atausw\ and (a
version of) \bbbb, we have
$${1\over2^{g-2}(g-2)!}\int_{\Sigma^{g-2}}\prod_k|dw_k|^2
\det \omega(z_1,z_2,w_{1},\ldots,w_{g-2})\det
\bar\omega(z_3,z_4,w_1,\ldots,w_{g-2})=
$$
$$\sum_{i<j\atop
m<n}(-1)^{i+j+m+n}[\omega_i(z_1)\omega_j(z_2)-\omega_j(z_1)
\omega_i(z_2)] \det_{k\neq i,j\atop l\neq m,n} {\rm
Im}\,\Omega_{kl}[\bar\omega_m(z_3)\bar\omega_n(z_4)-\bar
\omega_n(z_3)\bar\omega_m(z_4)] \ . $$ \vskip 6pt \noindent This
shows that partial integration on the variables of products of
determinants provides a method to construct linear combinations of
$\omega_i(z_1)\omega_j(z_2)\bar\omega_k(z_3)\bar\omega_l(z_4)$, with
the requested properties under modular transformations.

\subsec{The integration over $\Sigma^{2g}$}

In \DHokerHT, the amplitudes calculated using the genus two formula
were compared with the corresponding terms in the type IIB low
energy effective action. In particular, the two-loop contribution to
the $D^4R^4$ term, predicted by $SL(2,\ZZ)$ duality \Vanh, was
obtained by setting the momenta $k_i=0$ in the exponential of the
integrand of the 4-points amplitude formula. More precisely, in
\DHokerHT\ the identity
\eqn\dokid{\Delta(z_1,z_2)\Delta(z_3,z_4)+\Delta(z_1,z_4)\Delta(z_2,z_3)
+\Delta(z_1,z_3)\Delta(z_4,z_2)=0\ ,} where
$\Delta(z_i,z_j):=\omega_1(z_i)\omega_2(z_j)-\omega_2(z_i)\omega_1(z_j)$,
implied \eqn\dhokrel{\int_{\Sigma^4}|{\cal
Y}_S|^2=\int_{\Sigma^4}|s\Delta(z_1,z_4)\Delta(z_2,z_3)-t\Delta(z_1,z_2)
\Delta(z_3,z_4)|^2=32(s^2+t^2+u^2)(\det  \rm Im\,\Omega)^2 \ . }

We note that, at generic $g$, the symmetries of the function ${\cal
Y}_S$ under the exchange $(z_i,k_i)\leftrightarrow (z_j,k_j)$ and
the relation $s+t+u=0$ constrain $\int_{\Sigma^{2g}}|{\cal Y}_S|^2$
to be proportional to $(s^2+t^2+u^2)\det ({\rm Im\,}\Omega)^2$. The
term $s^2+t^2+u^2$ is the unique homogeneous polynomial quadratic in
the Mandelstam variables which is symmetric under the exchanges
$u\leftrightarrow s$, $t\leftrightarrow s$, $u\leftrightarrow t$.
Furthermore, since ${\cal Y}_S$ is a product of two determinants
made of the canonical basis of holomorphic one-differentials,
modular invariance implies that this integral must be proportional
to $(\det{\rm Im}\,\Omega)^2$.

In the following we will show that, by a suitable choice of the
coefficients $C(\rho)$ defining $\H_{ij,kl}$ in \Hijkl, the relation
\dokid\ can be generalized to

\eqn\nostid{\H_{12,34}+\H_{14,23}+\H_{13,42}=0\ .} The above remarks
suggest that with such a choice of the coefficients $C(\rho)$, a
generalization of \dhokrel\ holds also at higher genus. It is
instructive to explicitly perform the calculations that will also
provide the exact numerical factors.

\vskip 6pt

\noindent In the remaining of this section we will consider the
following steps.

\vskip 3pt

\item{1.} First we will derive a formula that will allow us to easily compute

$$\eqalign{{\cal L}_1 &:=\int_{\Sigma^{2g}}|\H_{12,34}|^2\ ,\cr\cr
{\cal L}_2 &:=\int_{\Sigma^{2g}}\H_{14,23}\bar{\H}_{12,34}\ .\cr}$$

\vskip 3pt

\item{2.} Next we will show that the condition \nostid\ is
equivalent to ${\cal L}_1=-2{\cal L}_2$ and this will be used to find
a class of coefficients $C(\rho)$ satisfying \nostid.

\item{3.} Finally, for this class of coefficients, we will
compute $\int_{\Sigma^{2g}}|{\cal Y}_S|^2$.

\subsec{Combinatorics of determinants}

Let $\tau$ be a permutation of the set $\{1,2,\ldots,2g\}$. We
define the following sets
$$\eqalign{D_1 &:=\{1,2,\ldots,g\}\cap\tau^{-1}(\{1,2,\ldots,g\})\ ,\cr\cr
D_2 &:=\{g+1,g+2,\ldots,2g\}\cap\tau^{-1}(\{g+1,g+2,\ldots,2g\})\
,\cr\cr I_1 &:=\{g+1,g+2,\ldots,2g\}\cap\tau^{-1}(\{1,2,\ldots,g\})\
,\cr\cr I_2 &:=\{1,2,\ldots,g\}\cap\tau^{-1}(\{g+1,g+2,\ldots,2g\})\
.}$$

\vskip 3pt

\noindent Note that ${\rm Card}\,D_1={\rm Card}\,D_2=g-m$, where
$m:={\rm Card}\,I_1={\rm Card}\,I_2$.

\vskip 3pt

\proclaim Theorem 3. Let $\{x_1,x_2,\ldots,x_{2g}\}$ be a set of
$2g$ points in $\Sigma$, $\tau$ a permutation of $\{1,2,\ldots,2g\}$
and $D_1,D_2,I_1,I_2$ and $m$ as above. Then
$$\eqalign{P(\tau) &:=\int_{\Sigma^{2g}} X(x_{1\rightarrow g})
X(x_{g+1\rightarrow 2g})\bar{X} (x_{\tau^{-1}(1\rightarrow
g)})\bar{X}(x_{\tau^{-1}(g+1\rightarrow 2g)})=\cr &=\pm
g!(g-m)!m!(\det A)^2\ ,}$$ where $A:= 2{\rm Im}\,\Omega$, and the
sign depends on the permutation $\tau$.

\noindent {\sl Proof:} We have
$$\eqalign{P(\tau) &=\int_{\Sigma^{2g}} \epsilon(i)\epsilon(j)
\epsilon(k)\epsilon(l)
 \omega_{i_1}(x_1)\ldots\omega_{i_g}(x_g)\,\omega_{j_1}(x_{g+1})
 \ldots\omega_{j_g}(x_{2g})\cdot\cr
 \cdot &\prod_{r\in D_1}\bar{\omega}_{k_{\tau(r)}}(x_r)
 \prod_{s\in I_1}\bar{\omega}_{k_{\tau(s)}}(x_s)
 \prod_{t\in I_2}\bar{\omega}_{l_{\tau(t)-g}}(x_t)
 \prod_{u\in D_2}\bar{\omega}_{l_{\tau(u)-g}}(x_u)\ ,}$$
where $\epsilon(i):= \epsilon^{i_1i_2\ldots i_g}$. Performing the
integration, we obtain
\eqn\Pexpr{P(\tau)=\epsilon(i)\epsilon(j)F(i,j)\ ,} where
$$F(i,j):=\epsilon(k)\epsilon(l)\prod_{r\in D_1}A_{i_rk_{\tau(r)}}
\prod_{s\in I_1}A_{j_{s-g}k_{\tau(s)}}\prod_{t\in
I_2}A_{i_tl_{\tau(t)-g}}\prod_{u\in D_2}A_{j_{u-g}l_{\tau(u)-g}}\
.$$ The indices $i$'s and $j$'s are just permutations of the set
$\{1,2,\ldots,g\}$, because of the factors $\epsilon(i)\epsilon(j)$
in \Pexpr. This means that, for each $s\in I_1$, $j_{s-g}$ must be
equal to one and only one $i_t$, $t\in D_1\cup I_2$. If
$i_{\hat{r}}=j_{\hat{s}-g}$ for some $\hat{r}\in D_1$ and
$\hat{s}\in I_1$, we can write
$$F(i,j)=\epsilon(k)\epsilon(l)A_{i_{\hat{r}}k_{\tau(\hat{r})}}
A_{j_{\hat{s}-g}k_{\tau(\hat{s})}}\cdot (\hbox{other terms})\ ,$$
where the terms in the brackets do not depend on $k_{\tau(\hat{r})}$
and $k_{\tau(\hat{s})}$. In this case, $F(i,j)$ is both symmetric
and skew-symmetric in $k_{\tau(\hat{r})}$ and $k_{\tau(\hat{s})}$,
so it must vanish identically. Thus, a necessary condition for
$F(i,j)$ to be non-vanishing is the existence of a bijective map
$$\eta:I_2\rightarrow I_1\ ,$$
such that $j_{\eta(t)-g}=i_t$ for all $t\in I_2$. In this case, we can write
$$F(i,j)
=\epsilon(k)\epsilon(l) \prod_{r\in
D_1}A_{i_rk_{\tau(r)}}\prod_{t\in I_2} A_{i_tk_{\tau\circ\eta(t)}}
\prod_{s\in I_1}A_{j_{s-g}l_{(\tau\circ\eta^{-1})(s)-g}}\prod_{u\in
D_2}A_{j_{u-g}l_{\tau(u)-g}}\ .$$ Let us define
\eqn\ladef{\lambda(r):=\left\{\matrix{\tau(r)\hfill &{\rm if\ }r\in
D_1\ ,\hfill \cr \tau\circ\eta\,(r) \hfill &{\rm if\ }r\in I_2\
,\hfill}\right.} and \eqn\mudef{\mu(t):=\left\{\matrix{\tau(t)\hfill
&{\rm if\ }t\in D_2\ ,\hfill \cr \tau\circ\eta^{-1}(t) \hfill &{\rm
if\ }t\in I_1\ ,\hfill}\right.} which are permutations of the sets
$\{1,\ldots,g\}$ and $\{g+1, \ldots,2g\}$, respectively. We have
\eqn\Fval{\eqalign{F(i,j)
&=\epsilon(k)\epsilon(l)\prod_{r=1}^{g}A_{i_rk_{\lambda(r)}}
\prod_{s=1}^{g}A_{j_sl_{\mu(s+g)-g}}=\cr
&=\epsilon(k)\epsilon(l){\rm sgn}\,(\lambda){\rm sgn}\,(\mu)
\prod_{r=1}^{g}A_{i_rk_r}\prod_{s=1}^{g}A_{j_sl_s}=\epsilon(i)\epsilon(j){\rm
sgn}\, (\lambda){\rm sgn}\,(\mu)(\det A)^2\ ,}} where we have
relabeled $k_{\lambda(r)}\rightarrow k_r$ and
$l_{\mu(s+g)-g}\rightarrow l_s$. The term ${\rm sgn}\,(\lambda){\rm
sgn}\,(\mu)$ comes from the rearrangement of the $k$'s and $l$'s in
$\epsilon(k)\epsilon(l)$, and we used the identity
$$\epsilon(k)\prod_{r=1}^gA_{i_rk_r}=\epsilon(i)\det A\ .$$
Note that, for a fixed $\tau$, there are, in general, several
different choices for the map $\eta$. If we change
$\eta\rightarrow\eta'$, we have
$$\eqalign{&{\rm sgn}\,(\lambda')={\rm sgn}\,
(\eta'\circ\eta^{-1})\,{\rm sgn}\,(\lambda)\ ,\cr &{\rm
sgn}\,(\mu')={\rm sgn}\,(\eta'{}^{-1}\circ\eta)\,{\rm sgn}\,(\mu)\
,}$$ and then
$${\rm sgn}\,(\lambda'){\rm sgn}\,(\mu')={\rm sgn}\,(\lambda){\rm sgn}\,(\mu)\ .$$
Therefore, the product ${\rm sgn}\,(\lambda){\rm sgn}\,(\mu)$
depends only on $\tau$ and not on $\eta$. We find that if $F(i,j)$
is nonzero, then it depends on the $i$'s and $j$'s only through the
product $\epsilon(i)\epsilon(j)$.

We now insert \Fval\ into \Pexpr\ and sum over all the sets of
indices $i$ and $j$ for which $F(i,j)$ is not null. We can
arbitrarily choose the values of $i_1,\ldots,i_g$; there are $g!$
different possibilities. We must choose the $j$'s in such a way that
there exists a function $\eta$ as above; this is a constraint on the
values of $j_{s-g}$, $s\in I_1$. There are exactly $m!$ different
bijections $\eta$ from $I_2$ to $I_1$. Finally we can choose the
values of the remaining $g-m$ indices $j_u$, for $u\in D_2$, that
gives a factor $(g-m)!$. We obtain
$$P(\tau)={\rm sgn}\,(\lambda){\rm sgn}\,(\mu)g!(g-m)!m!(\det A)^2\ .$$

\subsec{Evaluation of ${\cal L}_1$ and ${\cal L}_2$.}

We now use theorem 3 to compute ${\cal L}_1$ and ${\cal L}_2$. These
terms split into two contributions
$${\cal L}_i=
2({\cal L}_i^A+{\cal L}_i^B)\ ,
$$
where
\eqn\calll{{\cal L}_i^{A,B}=
\sum_{\rho,\sigma\in B'_{2g-4}}
C(\rho)\bar{C}(\sigma)\int_{\Sigma^{2g}}L_i^{A,B}\ , } and
$$L_1^A:=X_{12}(w_{\rho(1\rightarrow g-2)})X_{34}
 (w_{\rho(g-1\rightarrow 2g-4)})\bar{X}_{12}
 (w_{\sigma(1\rightarrow g-2)})\bar{X}_{34}
 (w_{\sigma(g-1\rightarrow 2g-4)})\ ,$$
$$L_1^B:=X_{12}(w_{\rho(1\rightarrow g-2)})X_{34}
(w_{\rho(g-1\rightarrow 2g-4)})\bar{X}_{34} (w_{\sigma(1\rightarrow
g-2)})\bar{X}_{12}(w_{\sigma(g-1\rightarrow 2g-4)})\ ,$$
$$L_2^A:=X_{14}(w_{\rho(1\rightarrow g-2)})X_{23}
 (w_{\rho(g-1\rightarrow 2g-4)})\bar{X}_{12}
 (w_{\sigma(1\rightarrow g-2)})\bar{X}_{34}
 (w_{\sigma(g-1\rightarrow 2g-4)})\ ,$$
$$L_2^B:=X_{14}(w_{\rho(1\rightarrow g-2)})X_{23} (w_{\rho(g-1\rightarrow
2g-4)})\bar{X}_{34}(w_{\sigma(1\rightarrow
g-2)})\bar{X}_{12}(w_{\sigma(g-1\rightarrow 2g-4)})\ .$$ Let us
consider ${\cal L}_1^A$. By the change of variables
$w_{\rho(i)}\rightarrow w_i$ we can replace $L_1^A$ in \calll\ by
$$X_{12}(w_{1\rightarrow
g-2})X_{34}(w_{g-1\rightarrow 2g-4})\bar{X}_{12}
(w_{\rho^{-1}\circ\sigma(1\rightarrow g-2)})
\bar{X}_{34}(w_{\rho^{-1}\circ\sigma(g-1\rightarrow 2g-4)})\ .$$ We
define the variables $\{x_i\}$, $i=1,\ldots,2g$ as
$$\matrix{x_1:=z_1\ ,\hfill &\qquad\hskip 3cm &x_{g+1}:=z_3\
,\hfill \cr x_2:=z_2\ , \hfill & &x_{g+2}:=z_4\ ,\hfill\cr
x_i:=w_{i-2}\ ,\hfill &i=3,\ldots,g\ ,\hfill
 &x_i:=w_{i-4}\ ,\hfill &i=g+3,\ldots,2g\ .\hfill \cr}$$
Then, we have
$${\cal L}_1^A=\sum_{\rho,\sigma\in B'_{2g-4}}C(\rho)
\bar{C}(\sigma)P(\tau)\ ,$$ where $P(\tau)$ is the expression in the
previous theorem and $\tau$ is the permutation
\eqn\detau{\matrix{\tau(1)=1\ ,\hfill &\qquad \hskip 2cm
&\tau(g+1)=g+1\ ,\hfill \cr \tau(2)=2\ , \hfill & &\tau(g+2)=g+2\
,\hfill \cr \tau(i)=\sigma^{-1}\circ\rho(i-2)\ ,\hfill
&i=3,\ldots,g\ ,\hfill &\tau(i)=\sigma^{-1}\circ\rho(i-4)\
 ,\hfill &i=g+3,\ldots,2g\ .\hfill \cr}}
Let us define \eqn\inters{n:={\rm Card}
\left[\rho(\{1,2,\ldots,g-2\})\cap\sigma(\{g-1,g,
\ldots,2g-4\})\right]\ ,} and notice
that $$n={\rm Card}[\{3,4,\ldots,g\}
\cap\tau^{-1}(\{g+3,\ldots,2g\})]\ .$$ Since $\{1,2\}\subset D_1$, and
$\tau^{-1}(\{g+1,g+2\})\subset D_2$, the above intersection corresponds
 to the set $I_2$. Then, by theorem 3 with $m=n$ we obtain
$${\cal L}_1^A=\sum_{n=0}^{g-3}F_ng!(g-n)!n!(\det A)^2\ ,$$
where \eqn\Fn{F_n=\sum_{(\rho,\sigma)\in B'_{2g-4}(n)}C(\rho)
\bar{C}(\sigma)S(\rho,\sigma)\ .} Here the sum is over all the pairs
$(\rho,\sigma)\in B'_{2g-4}(n)\subset B'_{2g-4}\times B'_{2g-4}$,
such that the intersection defined in \inters\ has $n$ elements, and
$$S(\rho,\sigma)={\rm sgn}\, (\lambda){\rm sgn}\,(\mu)\ ,$$ where
$\lambda$ and $\mu$ are defined in \ladef\ and \mudef, with $\eta$
an arbitrary bijective function from $I_2$ into $I_1$. Note that
$\lambda(1)=1$, $\lambda(2)=2$, $\mu(g+1)=g+1$ and $\mu(g+2)=g+2$.

\vskip 10pt

\noindent A similar calculation can be performed for ${\cal L}_1^B$.
We use the same change of variables for the $\{w_i\}$ and the same
definition for the variables $\{x_i\}$. In this case, the
permutation $\tau$ is given by
$$\matrix{\tau(1)=g+1\ ,\hfill \hskip 2cm &\tau(g+1)=1\ ,\hfill \cr
 \tau(2)=g+2\ , \hfill &\tau(g+2)=2\ ,\hfill
}$$ while the other entries are the same as in the previous case. We
define $n$ as in \inters. In this case
$\{1,2\}=\tau^{-1}(\{g+1,g+2\})\subset I_2$, so that, we can apply
theorem 3 with $m=n+2$ to get
$${\cal L}_1^B=\sum_{n=0}^{g-3}F_ng!(g-n-2)!(n+2)!(\det A)^2\ ,$$
where $F_n$ is defined in \Fn. No extra minus sign comes from the
product ${\rm sgn}\,(\lambda){\rm sgn}\,(\mu)$. To see this, notice
that we can choose an arbitrary $\eta$ in the definition, since this
product does not depend on this choice. We set $\eta(1)=g+1$ and
$\eta(2)=g+2$, and obtain $\lambda(1)=1$, $\lambda(2)=2$,
$\mu(g+1)=g+1$ and $\mu(g+2)=g+2$. These are exactly the same
relations as for ${\cal L}_1^A$, and then we get the same overall
sign.

\vskip 10pt

\noindent Let us now consider ${\cal L}_2^A$ and ${\cal L}_2^B$. In
the case of ${\cal L}_2^A$ the permutation $\tau$ is defined by
$$\matrix{\tau(1)=1\ ,\hfill &\tau(g+1)=2\ ,\hfill\cr
\tau(2)=g+2\ , \hfill\hskip 2cm &\tau(g+2)=g+1\ ,\hfill}$$ while in
the case of ${\cal L}_2^B$ we have
$$\matrix{\tau(1)=g+1\ ,\hfill\hskip 2cm &\tau(g+1)=g+2\ ,\hfill\cr
\tau(2)=2\ , \hfill &\tau(g+2)=1\ .\hfill}$$ We have $\{1\}\subset
D_1$, $\{2\}\subset I_2$, $\{g+1\}\subset I_1$ and $\{g+2\}\subset
D_2$ in the ${\cal L}_2^A$ case, while $\{1\}\subset I_2$,
$\{2\}\subset D_1$, $\{g+1\}\subset D_2$ and $\{g+2\}\subset I_1$ in
the ${\cal L}_2^B$ case. Thus, we set $m=n+1$ and obtain
$${\cal L}_2^A={\cal L}_2^B=-\sum_{n=0}^{g-3}F_ng!(g-n-1)!
(n+1)!(\det A)^2\ .$$The minus sign arises because of the product
${\rm sgn}\,(\lambda){\rm sgn}\,(\mu)$. To see this in the ${\cal
L}_2^A$ case, we choose $\eta(2)=g+1$ and obtain $\lambda(1)=1$,
$\lambda(2)=2$, $\mu(g+1)=g+2$ and $\mu(g+2)=g+1$. The interchange
$g+1\leftrightarrow g+2$ in the $\mu$ permutation gives the minus
sign. A similar computation can be performed for ${\cal L}_1^B$.

\vskip 10pt \noindent Summarizing, we have
$$\eqalign{{\cal L}_1 &=2\sum_{n=0}^{g-3}
F_ng![(g-n)!n!+(g-n-2)!(n+2)!]\,2^{2g}(\det {\rm Im}\,\Omega)^2\
,\cr {\cal L}_2 &=-4\sum_{n=0}^{g-3}F_ng!(g-n-1)!(n+1)!\,2^{2g}(\det
{\rm Im}\,\Omega)^2\ ,}$$ where $F_n$ is defined in \Fn. We remark
that these formulas hold for generic choices of the coefficients
$C(\rho)$.

\subsec{A condition on the building blocks}

We claimed that the condition \nostid\ is equivalent to the
following constraint \eqn\constr{{\cal L}_1=-2{\cal L}_2\ ,} that is
\eqn\Frel{\sum_{n=0}^{g-3}F_n[(g-n)!n!+(g-n-2)!(n+2)!-4(g-n-1)!(n+1)!]=0\
.}

\noindent Actually, we trivially have
$$\eqalign{&3{\cal L}_1=\int_{\Sigma^{2g}}|\H_{12,34}|^2+
\int_{\Sigma^{2g}}|\H_{13,42}|^2+\int_{\Sigma^{2g}} |\H_{14,23}|^2\
,\cr &3{\cal L}_2=\int_{\Sigma^{2g}}\H_{12,34}\bar\H_{13,42}+
\int_{\Sigma^{2g}}\H_{12,34}\bar\H_{14,23}+
\int_{\Sigma^{2g}}\H_{13,42}\bar\H_{14,23}\ ,}$$ where, in each sum,
the three terms are related each other by suitable changes of
variables. Then, the relation $3({\cal L}_1+2{\cal L}_2)=0$ reads
$$\int_{\Sigma^{2g}}|\H_{12,34}+\H_{14,23}+\H_{13,42}|^2=0\ ,$$ that
is Eq.\nostid. On the other hand, if \nostid\ holds, then
$$\int_{\Sigma^{2g}}|{\cal Y}_S|^2=\int_{\Sigma^{2g}}
|s\H_{14,23}-t\H_{12,34}|^2=(s^2+t^2){\cal L}_1-2st{\cal L}_2\ .$$
Since $\int_{\Sigma^{2g}}|{\cal Y}_S|^2$ is proportional to
$s^2+t^2+u^2=2(s^2+t^2+st)$, this implies \constr. Note that we can
use the formulas for ${\cal L}_1$ and ${\cal L}_2$ to evaluate
$\int_{\Sigma^{2g}}|{\cal Y}_S|^2$.

\vskip 10pt

In the following we will consider the general solutions for the
constraints \Frel\ in the cases $g=3$ and $g=4$ and then we will
propose a simple solution for general $g$. In this case, we will
also compute the integral $\int_{\Sigma^{2g}}|{\cal Y}_S|^2$.

\vskip 10pt

\noindent In the case $g=3$, there are two additional points $w_1$
and $w_2$. There is only a permutation in $B'_{2g-4}$; its
coefficient $C$ is just a normalization of $\H_{ij,kl}$. We have
$F_0=|C|^2$ and the condition \Frel\ is an identity.

\vskip 10pt

\noindent For $g=4$, we have four additional points $w_1,\ldots,w_4$
and three different permutations in $B'_{2g-4}$
$$\rho_1=(1,2,3,4)\ , \quad \rho_2=(1,4,3,2)\ ,
\quad \rho_3=(1,3,2,4)\ ,$$ with coefficients $C_1,C_2,C_3$,
respectively. Computing the signs $S(\rho_i,\rho_j)$, $i,j=1,2,3$,
we find $$\eqalign{F_0 &=|C_1|^2+|C_2|^2+|C_3|^2\ , \cr F_1
&=2\,{\rm Re}\,(C_1\bar{C}_2-C_2\bar{C}_3+C_3\bar{C}_1)\ ,}$$ and
the condition \Frel\ gives $$|C_1-C_2-C_3|^2=0\ .$$ Thus, we have
infinite real or complex solutions for the coefficients
$C_1,C_2,C_3$.

\vskip 10pt

\noindent For general $g$, there always exists a rather simple
solution to \Frel. Let us denote by $S'_{2g-4}\subset B'_{2g-4}$ the
set of all permutations $\rho$ such that
\eqn\permuta{\rho(i)=\left\{\matrix{1\hfill &\quad i=1\ ,\hfill \cr
i\hbox{ or }i+g-1\hfill &\quad i=2,\ldots, g-2\ ,\hfill\cr g-1\hfill
&\quad i=g-1\ ,\hfill \cr i\hbox{ or }i-g+1\hfill &\quad
i=g,\ldots,2g-4\ .\hfill}\right.} At genus $g$, this set has
$2^{g-3}$ elements. Then, set $$C(\rho)= \left\{\matrix{1\hfill\quad
&\rho\in S'_{2g-4}\ ,\hfill\cr\cr 0\hfill\quad &\hbox{otherwise}\
,\hfill}\right.$$ where we absorbed a possible overall constant in
$B_g$ in \aquattro. In the following we show that this is a
solution to \Frel.

Fix a pair $(\rho,\sigma)\in S'_{2g-4}$, and define $\tau$ as in
\detau. Each element $r\in I_2$ is given by $r=\tau^{-1}(r+g-1)$.
Thus, we can choose $\eta(r)=r+g-1$, and we have $\lambda=\mu={\rm
id}$. This means that $S(\rho,\sigma)=+1$ for all $\rho,\sigma\in
S'_{2g-4}$.

Let us now compute $F_n$. Since $C(\rho)\bar
C(\sigma)S(\rho,\sigma)=1$ for all $\rho,\sigma\in S'_{2g-4}$, it
follows that $F_n$ is just the number of pairs $(\rho,\sigma)$
satisfying \inters. We can choose $\rho$ arbitrarily; the
permutations $\sigma$ satisfying \inters\ are in one to one
correspondence with the ways of choosing $n$ elements in the set
$\{2,\ldots,g-2\}$. Thus $$F_n=2^{g-3}{(g-3)!\over (g-n-3)!n!}\ ,$$
and by \Frel, we obtain the condition
\eqn\srel{{\sum_{n=0}^{g-3}(g-n-2)[(g-n)(g-n-1)+(n+1)(n+2)
-4(g-n-1)(n+1)]=0\ .}} Now observe that the identity
\eqn\sempl{\sum_{n=0}^{g-3}(g-n-2)(n+2)(n+1)=\sum_{n=0}^{g-3}(n+1)(g-n-1)(g-n-2)\
,} which follows by a redefinition $n\rightarrow g-n-3$, also
implies
$${\sum_{n=0}^{g-3}(g-n-2)[(g-n)(g-n-1)+(n+1)(n+2)]
=(g+1)\sum_{n=0}^{g-3}(g-n-1)(g-n-2)\ ,}$$
and
$$4\sum_{n=0}^{g-3}(n+1)(g-n-1)(g-n-2)=2(g+1)
\sum_{n=0}^{g-3}(n+1)(g-n-2)\ .$$
Then, \srel\ reduces to
$$\sum_{n=0}^{g-3}(g-3n-3)(g-n-2)=0\ ,$$
which is an identity, as one can verify using
\eqn\qsum{\sum_{n=0}^{g-3}n^2={1\over6}(g-2)[2(g-2)^2-3(g-3)-2]\ .}

\vskip 10pt

\noindent Thus, we have found a simple form for the building block
\eqn\bblock{\H_{ij,kl}=\sum_{\rho\in
S'_{2g-4}}[X_{ij}(w_{\rho(1\rightarrow
g-2)})X_{kl}(w_{\rho(g-1\rightarrow
2g-4)})+X_{kl}(w_{\rho(1\rightarrow
g-2)})X_{ij}(w_{\rho(g-1\rightarrow 2g-4)})]\ ,} where $S'_{2g-4}$
is the subgroup of the permutations $\rho$ satisfying \permuta. By
\constr\ and the kinematic relation $s+t+u=0$, we have
$$\int_{\Sigma^{2g}}|{\cal Y}_S|^2=-(s^2+t^2+u^2){\cal L}_2\ .$$ Eq.\bblock\
gives
$$-{\cal L}_2=2^{3g-1}g!(g-3)!\sum_{n=0}^{g-3}(g-n-1)
(g-n-2)(n+1)(\det {\rm Im}\Omega)^2\ ,$$ and by \sempl\ and \qsum,
we get

$$\int_{\Sigma^{2g}}|{\cal Y}_S|^2={2^{3g-3}\over 3}
(g+1)!(g-2)!(g^2-7g+18)(s^2+t^2+u^2)(\det{\rm Im}\,\Omega)^2\ .$$

\newsec{Conclusions.}

The 2-loop 4-point superstring amplitude, in the slice independent
formulation \phong, shows an unexpectedly simple shape and suggests
some possible generalization to higher genus. A natural conjecture
is that the measure on the moduli space is (the restriction of) the
modular invariant measure on the Siegel upper half-space derived in
\MatoneBX. Furthermore, we argued that the basic building block
generalizing the terms  $\Delta(z_i,z_j)\Delta(z_k,z_l)$ in the
genus 2 formula, is a sum of terms made of product of 2 determinants
of holomorphic abelian differentials. So that, we are naturally lead
to the following formula

\vskip 6pt

$$ A_4^{g{\rm-loop}} = B_g
\, \int_{\M_g}|\det g|^{1/2}{\rm d}\,\Xi_1\wedge\ldots\wedge {\rm
d}\,\bar\Xi_{3g-3} {\cal F}_g(k_i)\ .$$

\vskip 6pt

\noindent In the case of non-hyperelliptic Riemann surfaces of genus
$3$ we have

$$A_4^{3{\rm -loop}}=B_3\int_{\M_3}{|\bigwedge_{i\le j}^3 {\rm d}
\Omega_{ij}|^2\over (\det  {\rm Im} \,
\Omega)^4}\int_{\Sigma^6}{|{\cal Y}_S|^2\over(\det {\rm Im} \,
\Omega)^2 } \, e^{-\sum_{i<j}k_i\cdot k_j\,G(z_i,z_j)}\ ,$$ where,
since there are only two extra points, there is only a possible
choice for $\H_{ij,kl}$, namely
$$\H_{ij,kl}=\det\omega(z_i,z_j,w_1)\det\omega(z_k,z_l,w_2)
+\det\omega(z_i,z_j,w_2)\det\omega(z_k,z_l,w_1)\ .$$

\noindent In the case of genus $g\ge 4$ the definition of
$\H_{ij,kl}$ requires choosing a set of permutations over the $2g-4$
points $\{w_k\}$ and their relative coefficients. The simplest
choice is
$$\H_{12,34}=X_{12}(w_{1\rightarrow g-2})X_{34}(w_{g-1\rightarrow
2g-4})+ X_{34}(w_{1\rightarrow g-2})X_{12}(w_{g-1\rightarrow 2g-4})\
.$$

In the paper we considered the simplest generalization of \tauswa.
However, more general formulas for the building blocks $\H_{ij,kl}$
can be constructed in terms of products of $4$ determinants like
$\det\omega_i(x_j)$, times suitable functions of the kinematical
variables. In the four matrices $\omega_i(x_j)$, $j=1,\ldots,4g$
each point $x_j$ corresponds either to one of the insertion points
$z_1,\ldots,z_4$, or to one of the $4g-4$ additional points
$\{w_i\}$, to be eventually integrated away. The integration may
give factors that cancel some power in the overall $(\det {\rm Im}
\, \Omega)^4$ in the denominator inserted to balance the modular
transformations of the four determinants and their complex
conjugated. Thus, the basic step in defining such building blocks is
the distribution of the points $z_1,\ldots,z_4$ and $\{w_i\}$ in
such matrices. There are several different ways to implement such a
distribution. In particular, with respect to the position of the
points $z_1,\ldots,z_4$, in the determinants, there are five
possibilities \eqn\possibile{\eqalign{&\det\omega(z_i,w\ldots)\det
\omega(z_j,w\ldots)\det \omega(z_k,w\ldots)\det\omega(z_l,w\ldots)\
,\cr\cr &\det\omega(z_i,z_j,w\ldots)\det \omega(z_k,w\ldots)\det
\omega(z_l,w\ldots)\det\omega(w\ldots)\ ,\cr\cr
&\det\omega(z_i,z_j,w\ldots)\det \omega(z_k,z_l,w\ldots)\det
\omega(w\ldots)\det\omega(w\ldots)\ ,\cr\cr
&\det\omega(z_i,z_j,z_k,w\ldots)\det \omega(z_l,w\ldots)\det
\omega(w\ldots)\det\omega(w\ldots)\ ,\cr\cr
&\det\omega(z_i,z_j,z_k,z_l,w\ldots)\det \omega(w\ldots)\det
\omega(w\ldots)\det\omega(w\ldots)\ ,\cr }} \noindent where
$w\ldots$ is a permutation of a subset of suitable cardinality of
$\{w_i\}$. Note that the last two combinations of determinants may
appear only at $g\geq 3$ and $g\geq 4$ respectively. In the case of
determinants which do not contain any insertion point $z_i$, the
integration over the additional points leads to overall factors
$\det{\rm Im} \,\Omega$ reducing the power of  $(\det{\rm Im}
\,\Omega)^{4}$ in the denominator.

In some cases, this mechanism leads to formulas involving the
product of two determinants, as considered in this paper. However,
in principle, there could be some interesting generalizations which
require combinatorial analysis of rapidly growing complexity.

Remarkably, the construction suggests a natural generalization to
the $n$ point functions, where the products of $n$ determinants may
still play the role of building blocks. On the other hand, at the
moment we can hardly derive the exact kinematical factors they are
associated to. An extension of our analysis and the symmetry
properties of the amplitudes, could, in principle, constrain the
possible shape of such functions. We also note that the systematic
use of the determinants of holomorphic one-differentials, may lead
to control the factorization properties in the various degeneration
limits. This would suggest the appearance of an integrable structure
related to the cohomology of $\bar{\cal M}_g$.

\vskip 20pt

\noindent {\bf Acknowledgements}. We would like to thank G.
Bertoldi, G. Bonelli, L. Mazzucato, Y. Nakayama, P. Pasti, D.
Sorokin, M. Tonin, G. Travaglini, P. Vanhove and C.-J. Zhu, for
stimulating discussions. Work partially supported by the European
Community's Human Potential Programme under contract
MRTN-CT-2004-005104 ``Constituents, Fundamental Forces and
Symmetries of the Universe".

\listrefs

\end